\documentstyle[epsfig]{aipproc}

\begin{document}
\title{The Extragalactic Diffuse Gamma-Ray Emission}

\author{P. Sreekumar$^*$\footnote{USRA Research Scientist}
, F.W. Stecker$^*$ and S.C. Kappadath$^{**}$}
\address{$^*$NASA/Goddard Space Flight Center, Greenbelt, MD 20771\\
$^{**}$University of New Hampshire, Durham, NH 03824}

\maketitle

\def\etal{{et al.\,}}
\def\flux{{$\times$10$^{-5}\rm photons $ cm$^{-2}$s$^{-1}$ sr$^{-1}$}}
\def\deg{{$^{\circ}$}}
\begin{abstract}
The all-sky surveys in $\gamma$-rays by the imaging Compton telescope (COMPTEL) 
and the Energetic Gamma Ray Experiment Telescope (EGRET) on board the
Compton Gamma Ray Observatory for the first time allows detailed studies of the extragalactic diffuse emission at $\gamma$-ray 
energies greater 1 MeV.  A preliminary analysis of COMPTEL data 
indicates a significant decrease in the level of the derived 
cosmic diffuse emission from previous estimates in the 1--30 MeV range,
with no evidence for an MeV-excess, at least not at the levels
claimed previously. The 1--30 MeV flux measurements are compatible 
with power-law extrapolation from lower and higher energies.
These new results indicate that the possible contributions to the 
extragalactic emission from processes that could explain the MeV-excess,
such as matter-antimatter annihilation, is significantly reduced.
At high energies ($>$ 30 MeV), the
extragalactic emission is well described by a power law photon spectrum
with an index of --(2.10$\pm$0.03) in the 30 MeV to 100 GeV energy
range. No large scale spatial anisotropy or changes in the energy
spectrum are observed in the deduced extragalactic emission.
The most likely explanation for the origin
of this extragalactic $\gamma$-ray emission above 10 MeV, is that it 
arises primarily from unresolved $\gamma$-ray-emitting blazars. 
The consistency of the average $\gamma$-ray blazar spectrum with the 
derived extragalactic
diffuse spectrum strongly argues in favor of such an origin.
The extension of the power law spectrum to 100 GeV implies
the average emission from $\gamma$-ray blazars extends to 100 GeV.

\end{abstract}

\section*{Introduction}

The extragalactic diffuse emission at $\gamma$-ray energies
has interesting cosmological implications since the bulk of these
photons suffer little or no attenuation during their
propagation from the 
site of origin \cite{stecker73}.  
The first all-sky survey in low and medium energy $\gamma$-rays 
(1 MeV--30 MeV) has been carried out by COMPTEL and at higher 
energies ($>$30 MeV) by EGRET on board the Compton
Observatory satellite (CGRO). 
Improved sensitivity, low instrumental background and 
a large field of view of these instruments have resulted 
in significantly improved measurements of the
extragalactic $\gamma$-ray emission and our understanding 
of its origin.
The measurement of the extragalactic $\gamma$-ray emission is made
difficult by the very low intensity of the expected
emission and the lack of a spatial or temporal signature 
to separate the cosmic signal from other radiation. 
The measured diffuse emission along any line-of-sight, 
could be composed of a Galactic component arising from cosmic-ray
interactions with the local interstellar gas and radiation,
an instrumental background, and an extragalactic  component.
In the 1 to 10 MeV band, diffuse studies have been
traditionally complicated by difficulties in 
fully accounting for the instrumental background. 
At higher energies (E$>$10 MeV), the results are subject to 
difficulties in accurately accounting for the Galactic diffuse 
emission. 

Before the launch of CGRO,
the $\gamma$-ray spectrometer flown aboard
three Apollo flights \cite{trombka77} and numerous
balloon-borne experiments \cite{white77}\cite{schonfelder80},
showed the presence of a `bump' -like feature in the few MeV
range that was in excess of the extrapolated hard X-ray continuum. 
Although the measured intensities varied widely among the numerous
experiments, most showed some level of an MeV-excess. It was recognized
as early as 1972, that cosmic-ray induced radioactivity is the most
dominant source of background in the MeV energy range
\cite{fishman72}. 
Recent results from the COMPTEL \cite{kappadath96}\cite{kappadath97} and the
Solar Maximum Mission (SMM) \cite{watanabe97} experiments indicate no 
evidence for an MeV-bump, at least not at the levels previously reported.

At higher energies ($>$35 MeV), the SAS-2 satellite 
\cite{fichtel75} provided the first clear evidence for the 
existence of an extragalactic $\gamma$-ray component. 
This emission, seen as an excess over the the strong 
Galactic diffuse radiation, was uncorrelated with the column 
density of matter and was therefore interpreted as being extragalactic 
in origin \cite{fst78}.  The recent EGRET results 
\cite{sreekumar97} 
extend the high energy measurement to an unprecedented $\sim$100 GeV. 
The emission above 30 MeV is well represented by a single power-law of
index --2.1 and shows no significant  departure from isotropy.

The origin of the extragalactic $\gamma$-ray emission
has proved to be an elusive goal for theorists over the years. Prior to
CGRO, even the
question as to whether the radiation is from a truly diffuse process or is,
in fact, the superposition of radiation from a large number of extragalactic
sources has been difficult to answer. 
At MeV energies, theoretical efforts to explain the emission, were
constrained by the need to explain the MeV excess. 
The absence of a source class at these
energies, whose spectrum displayed this characteristic
signature, made this particularly difficult. At higher energies, the SAS-2 and COS-B experiments together,
detected one extragalactic source, 3C273, suggesting active
galactic nuclei (AGN) as a viable source class that could contribute to the
diffuse background \cite{bignami79}\cite{kazanas83}.
More recently, the detection of $>$50 $\gamma$-ray blazars by EGRET and their
spectral properties, have been used to improve theoretical
calculations of the diffuse $\gamma$-ray emission from blazars.

Here we review the current observational and theoretical understanding
of the diffuse extragalactic $\gamma$-ray emission above 1 MeV. 
Recent analysis results from the instruments on the 
Compton Gamma Ray Observatory are summarized. 
For a more detailed discussion on the COMPTEL and EGRET 
results, see Kappadath \etal (1996, 1997)\cite{kappadath96}\cite{kappadath97}
and Sreekumar \etal (1997)\cite{sreekumar97} respectively. 
Finally, the implications of these new findings on the origin of the 
extragalactic diffuse emission, are discussed.

\section*{Recent Results from CGRO}
\begin{figure}
\centerline{\epsfig{file=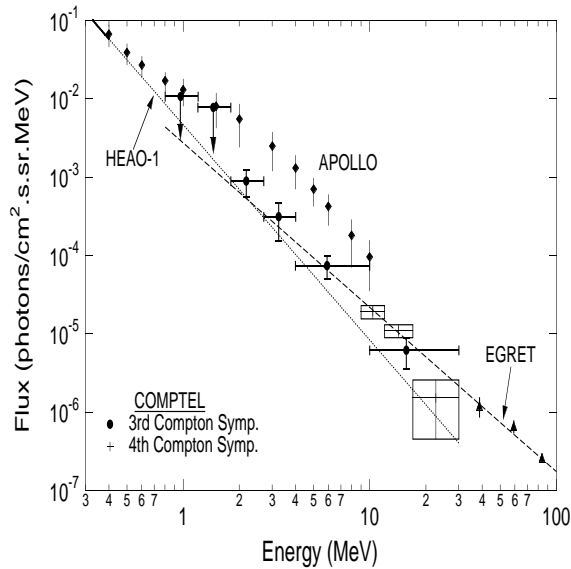,height=3.0in,width=3.0in}}
\vspace{10pt}
\caption{COMPTEL results (Kappadath \etal 1996, 1997): The HEAO-1 data 
are from Kinzer \etal (1997) and the Apollo measurements 
are from Trombka \etal (1977)}
\label{fig1}
\end{figure}

\begin{figure}
\centerline{\epsfig{file=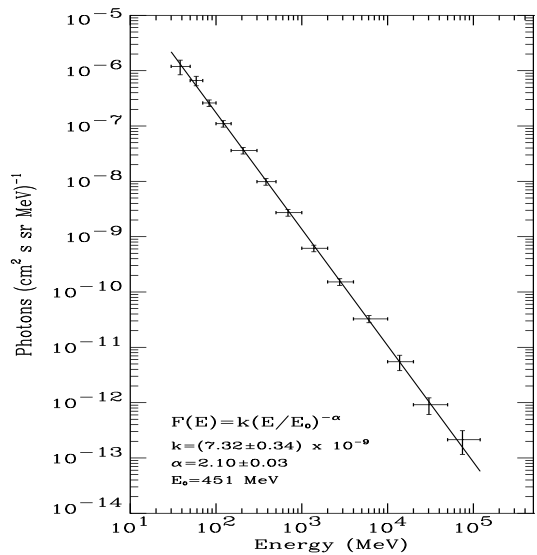,height=3.0in,width=3.0in}}
\vspace{10pt}
\caption{EGRET results: The extragalactic diffuse emission spectrum $>$30 MeV
(Sreekumar \etal 1997)}
\end{figure}

\begin{figure}[t]
\centerline{\epsfig{file=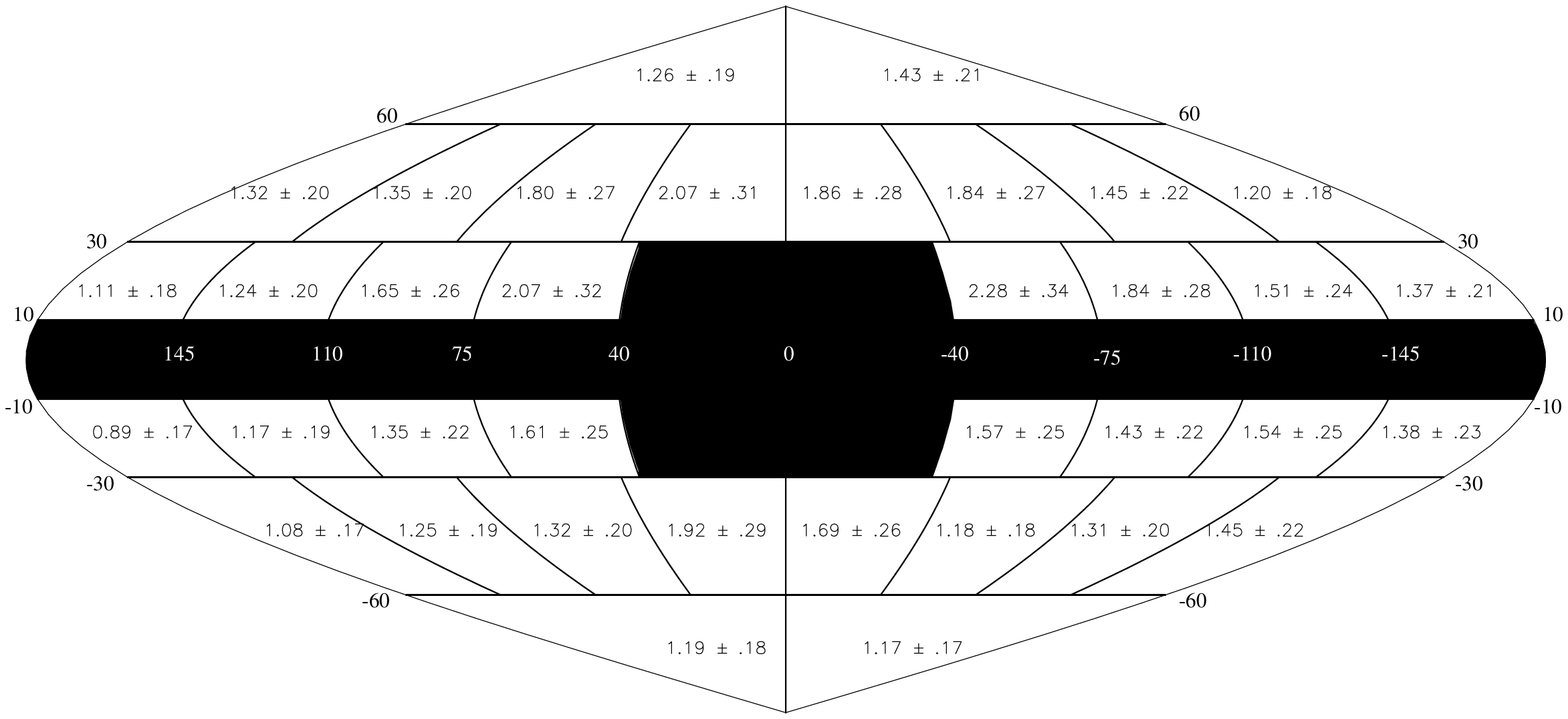,height=3.0in,width=5.0in,angle=-90}}
\vspace{10pt}
\caption{The distribution of extragalactic flux (E$>$100
MeV). The shaded region containing the Galactic plane is
excluded due to extreme dominance of the Galactic emission
and the region centered on the Galactic Center and extending 
towards $\pm$30\deg in latitude is excluded due
to difficulties in modeling all of the Galactic emission
(Sreekumar \etal 1997).}
\label{fig3}
\end{figure}

\subsection*{COMPTEL results (1--30 MeV)}
The COMPTEL diffuse emission spectrum is constructed from high-latitude
observations, by first subtracting the
instrument background and then attributing the residual flux to the
extragalactic diffuse radiation. The instrumental background is in
general composed of `prompt' and `long-lived' components.

The prompt background is instantaneously produced by proton and
neutron interactions in the spacecraft. Hence the prompt background
refers to the component that modulates with the instantaneous local
cosmic-ray flux. The prompt background is seen to vary linearly with
the veto-scalar-rates (charge-particle shield rates). 
Assuming that zero veto-scalar-rate corresponds to zero cosmic ray flux,
a linearly extrapolation is used to compute the prompt background
contribution.
  
The long-lived background events are due to de-excitation photons from
activated radioactive isotopes with long half-lives ($\tau_{1/2} >$ 30
sec). Their decay rate is not directly related to the instantaneous
cosmic-ray flux because of the long half-lives.  
The long-lived background
isotopes are identified by their characteristic decay lines in the
individual detector spectra. Monte Carlo simulation of the isotope
decay are used to determine the COMPTEL detector response. The
measured energy spectrum is used to determine the absolute
contribution of each of the long-lived background isotopes.
   
The diffuse flux measured by COMPTEL, refers to the total $\gamma$-ray flux in the
field-of-view ($\sim$1.5 sr) derived from high-latitude
observations. It is important to point out that it includes flux 
contribution from the Galactic diffuse
emission and $\gamma$-ray point sources in the field of view.     
It is important to point out that the COMPTEL results below $\sim$ 9 MeV
are still preliminary. The 2--9 MeV flux is significantly lower than
pre-COMPTEL measurements. They show no evidence for a
MeV-excess, at least not at the levels reported previously. Only
upper-limits are claimed below $\sim$2 MeV. 
Recent  improvements in the 9-30 MeV spectrum 
\cite{kappadath97}, are shown in figure 1. 
The 9--30 MeV flux is consistent with the previous measurements 
(mostly upper-limits) and also compatible with the
extrapolation of the EGRET spectrum \cite{sreekumar97} to lower energies.
     
In examining the isotropic nature of the diffuse radiation, a
simple comparison shows that the measured 9--30 MeV spectrum from the
Virgo and South Galactic Pole regions are consistent with each other
\cite{kappadath97}.

\subsection*{EGRET results (30 MeV -- 100 GeV)}

In the EGRET energy range, the primary source of error in estimating the 
extragalactic emission arises from uncertainties in the Galactic diffuse 
emission model.
In order to derive the extragalactic emission without being sensitive
to the Galactic model used, the following approach is adopted for the
EGRET data. 
The observed emission (I$_{observed}$)
is assumed to be made up of a Galactic (I$_{galactic}$)
and an extragalactic component (I$_{extragalactic}$).
 
\centerline{$I_{observed}$(l,b,E) = I$_{extragalactic} + B \times
I_{galactic}$(
l,b,E)}
 
\noindent The slope, `$B$' of a straight line fit to a plot of
observed emission versus the Galactic model gives an independent measure for
the normalization of the input Galactic model calculation.
The primary processes that
produce the observed Galactic diffuse $\gamma$-rays are:
cosmic-ray nucleons interacting with nucleons in the interstellar 
gas, bremsstrahlung by cosmic-ray electron, and inverse Compton 
interaction of cosmic-ray electrons with ambient low-energy 
interstellar photons \cite{stecker77}. 
The Galactic diffuse emission falls of rapidly at higher latitudes,
making high-latitude observations, ideally suited to study the
extragalactic emission.
The possible contribution to the Galactic diffuse emission from unresolved 
point sources such as pulsars,
is uncertain with estimates ranging from a few 
percent to almost 100$\%$ depending on the choice of many model parameters
such as the birth properties of 
pulsars \cite{bailes92}. The evidence for a pion `bump' ( from
neutral pion decay) in the Galactic diffuse spectrum 
\cite{hunter97} can be used to set an upper
limit for the contribution from unresolved sources at $<$50$\%$. 

\begin{figure}
\centerline{\epsfig{file=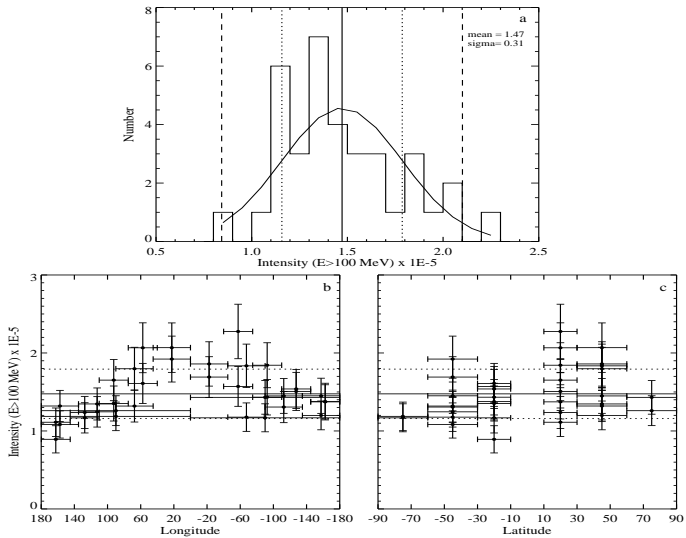,height=3.5in,width=4.0
in}}
\vspace{10pt}
\caption{The all-sky distribution of integral flux (E$>$100
MeV) in units of photons (cm$^2$-s-sr)$^{-1}$ (Sreekumar \etal 1997). 
(a) Gaussian fit to the flux histogram showing the mean (solid), 1$\sigma$
(dotted) and 2$\sigma$ (dashed) confidence intervals; (b)\&(c) Intensity 
values plotted over longitude and latitude respectively.  The longitude 
distribution shows slightly larger intensity values within $\pm$60 \deg}
\label{fig4}
\end{figure}
 
\begin{figure}[t]
\centerline{\epsfig{file=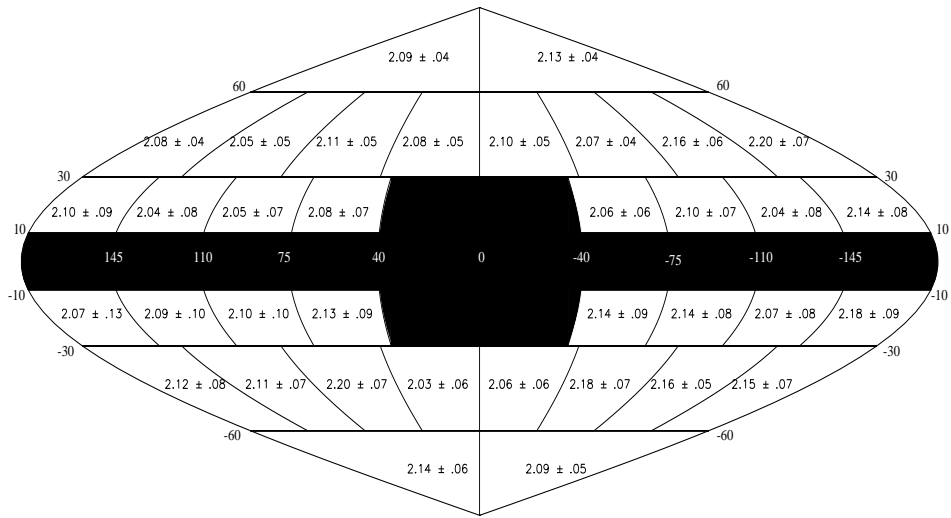,height=3in,width=5.0in,angle=-90}}
\vspace{10pt}
\caption{The distribution of spectral indices assuming a single power-law
fit to the extragalactic spectrum above 30 MeV (Sreekumar \etal 1997).}
\label{fig5}
\end{figure}
  
\begin{figure}[b]
\centerline{\epsfig{file=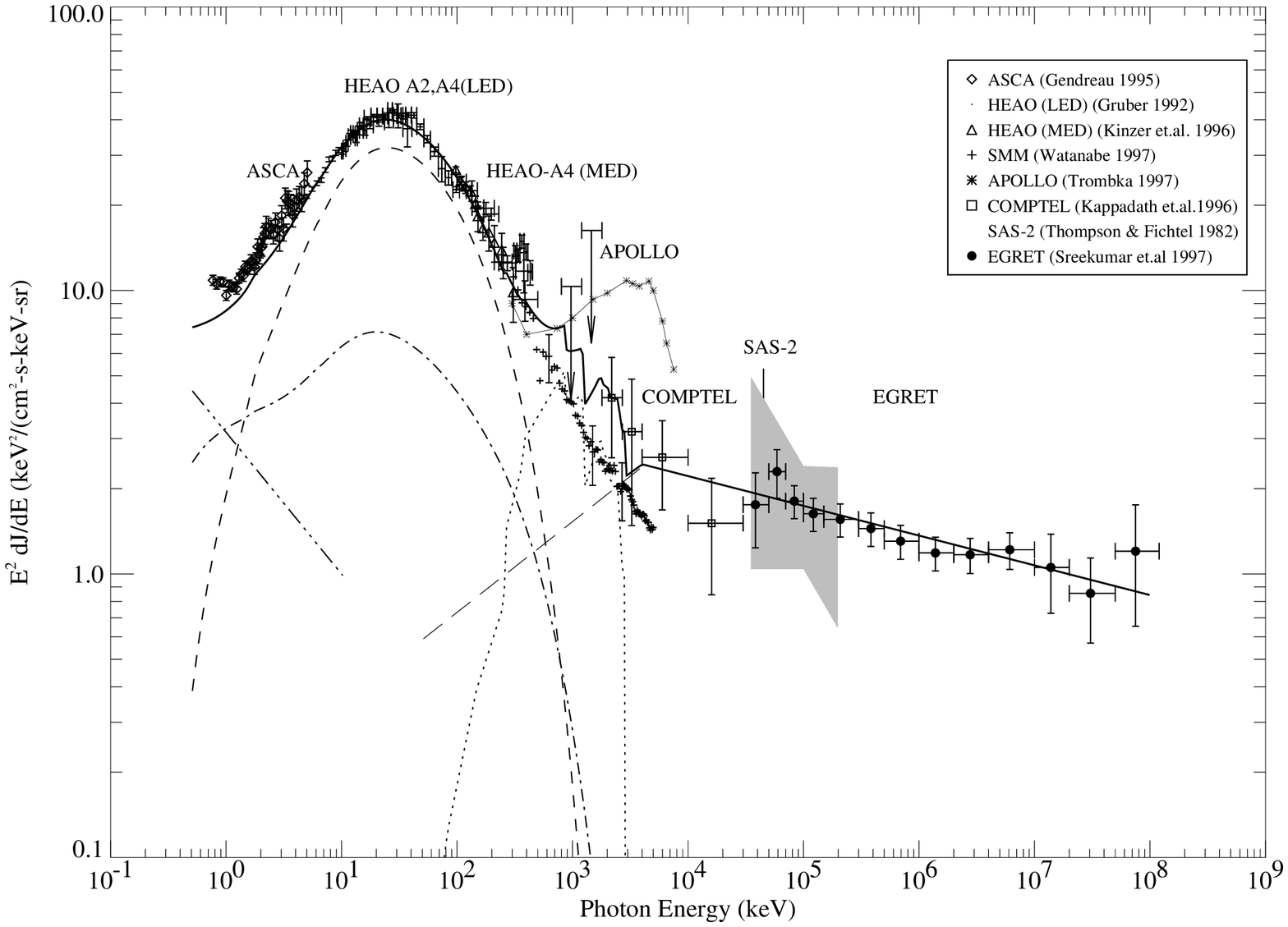,height=3.5in,width=6.0in,angle=90}}
\vspace{0pt}
\caption{Multiwavelength spectrum from X-rays to $\gamma$-rays
including the revised 1$\sigma$ upper limits from the Apollo experiment
(Trombka 1997).  The estimated contribution from Seyfert 1 
(dot-dashed), and Seyfert 2 (dashed) are from the model of Zdziarski(1996);
steep-spectrum quasar contribution (dot-dot-dashed) is taken from Chen \etal
(1997); Type Ia supernovae (dot) is from The, Leising and Clayton (1993). 
the average blazar spectrum breaks around 4 MeV (McNaron-Brown \etal 1995) 
to a power law with an index of $\sim$--1.7.  The thick solid line indicates 
the sum of all the components.}
\label{fig6}
\end{figure}

Preliminary results on the extragalactic spectrum above 30 MeV were reported by
Kniffen \etal (1996)\cite{kniffen96}, using the Galactic diffuse model of 
Hunter \etal (1997)\cite{hunter97} and its high-latitude extension discussed
by Sreekumar \etal (1997)\cite{sreekumar97}. Earlier, 
Osborne, Wolfendale and Zhang (1994)\cite{osborne94} had shown that the 
spectrum derived
from EGRET is well represented by a power-law of index (2.11$\pm$0.05).
Even though this was within errors of the previous best estimate of 
(2.35$^{+0.4}_{-0.3}$) from the SAS-2 experiment \cite{thompson82},
the EGRET measurements clearly demonstrated the existence of a well defined,
harder power-law spectrum. Independent analysis by Chen, Dwyer and Kaaret (1996)
\cite{cdk96}
also yielded a spectral index of (2.15$\pm$0.06)  
and an integral flux above 100 MeV of
(1.24$\pm$0.06)\flux for E$>$100 MeV.
More recently, Sreekumar \etal (1997)\cite{sreekumar97} using $\sim$ the first 4 years of 
EGRET observations, carried out a more detailed and careful analysis 
of the intensity and spectral shape in different regions covering the full 
sky and for the first time extended the spectrum up to an unprecedented 100 GeV.
The differential photon spectrum of
the extragalactic emission averaged over the sky is well fit
by a power law with an index of --(2.10$\pm$0.03). The spectrum was
determined using data from 30 MeV to 10 GeV; however as shown in Figure 2, the
differential photon flux in the 10 to 20 GeV, 20 to 50 GeV, and 50 to 120
GeV energy intervals are also consistent with extrapolation of the
single power law spectrum.  The integrated flux
from 30 to 100 MeV is (4.26$\pm$0.14)\flux and that above 100
MeV is (1.45$\pm$0.05)\flux.
 
With the availability of high quality data from CGRO, one can for the first 
time, address the isotropic nature of the extragalactic emission.
The derived integral fluxes
in 36 independent regions of the sky are shown in Figure
3, with the values ranging from a low of (0.89$\pm$0.17)\flux to a
high of (2.28$\pm$0.34)\flux \cite{sreekumar97}. The shaded area in the 
figure represents the region where the Galactic diffuse emission is dominant 
and hence not included in the analysis.
Figure 4a shows a histogram of integral flux values,
overlayed by a Gaussian fit assuming equal measurement errors.
The Gaussian fit yields a mean flux above 100 MeV of
1.47\flux with a standard deviation of 0.33\flux consistent 
with the all-sky calculation.
To further examine the spatial distribution, Figure 4b and
4c shows the same data plotted against Galactic longitude and
latitude respectively.
No significant deviation from uniformity
is observed in the latitude distribution; however as a
function of longitude,
there appears to be an enhancement in the derived intensities
towards l$\sim$20\deg, primarily due to a few
regions on the boundary of the excluded Galactic center region.
This could arise from unaccounted extended Galactic diffuse emission.
If one excludes the inner regions of the Galaxy ($|r| <$60\deg;
r=angle made with the direction of the Galactic center), the
mean flux is 1.36\flux above 100 MeV.
Thus the measured extragalactic flux is consistent within
errors to a uniform sky distribution in regions outside the inner Galaxy.
 
Figure 5 shows the distribution of the derived spectral indices from fits
to  data from 36 independent regions of the sky. The power law indices vary from
--(2.04$\pm$0.08) to --(2.20 $\pm$0.07) and show no systematic
deviations from the value of --(2.10$\pm$0.03), derived from
the all-sky analysis.  

\section*{Origin of the extragalactic $\gamma$-ray emission}
A large number of possible origins for the extragalactic diffuse $\gamma$-ray 
emission have been proposed over the years (e.g. \cite{stecker75}\cite{ft81}
\cite{gehrels96}). The key to understanding the origin of the extragalactic emission may 
lie in the realization that it may be composed of a number of different
components having different origins and that these components in turn may
dominate the observed emission only in specific energy ranges (Figure 6). For example, 
the putative matter-antimatter component would only be important in the energy 
range between $\sim 1$ MeV and $\sim 100$ MeV. The unresolved source
models must also be broken up into different source classes and energy ranges.
We discuss separately the two
distinct possibilities,  a truly diffuse origin or the superposition of
unresolved sources.

\subsection*{Diffuse origin}
The physics of the truly diffuse 
physical processes which might be involved have been discussed previously 
\cite{stecker71}\cite{stecker73}\cite{stecker75}.
Among those that could potentially contribute to the emission, 
particularly between 1 and 100 MeV, the most significant could be
matter-antimatter annihilation in a globally baryon-symmetric cosmology
based on grand unified theories and early universe physics \cite{stecker71b}
\cite{stecker85}.
In the light of recent COMPTEL and SMM
data, it appears more likely that the contribution from such a process (if
any) is significantly lower than previously reported.
Around $\sim$1 MeV, it has been suggested that a significant fraction 
of the emission could be made up of $\gamma$-ray production in type Ia 
and type II  supernovae primarily via lines from the decay of $^{56}$Ni 
$\rightarrow ^{56}$Co $\rightarrow ^{56}$Fe and from $^{26}$Al, $^{44}$Ti, 
and $^{60}$Co \cite{the93}\cite{watanabe97}.
However, uncertainties associated with the existing observational data 
in the 1--30 MeV makes it difficult to estimate the contributions 
from the different proposed processes.

Diffuse processes that have been proposed to explain the high-energy emission
include , primordial black hole evaporation \cite{page76}\cite{hawking77}, 
million solar-mass black holes which collapsed at high redshift 
(z$\sim$100) \cite{go92}, and some exotic source proposals, 
such as annihilation of supersymmetric particles
\cite{silk84}\cite{rudaz91}\cite{kam95}.
All of these theories predict continuum or line contributions 
that are unobservable above 30 MeV with EGRET. 

\subsection*{Unresolved Point Sources}
\subsubsection*{Normal Galaxies}
Models based on discrete source contributions have considered a variety of source classes.
Normal galaxies might at first appear to be a reasonable possibility for the origin of the 
diffuse radiation since they are known to emit $\gamma$-rays and to do so to very high 
$\gamma$-ray energies \cite{sreekumar92}\cite{hunter97}.  
Previous estimates 
\cite{kraushaar72}\cite{strong76}\cite{lichti78}\cite{fst78}
have shown
that the intensity above 100 MeV expected from normal galaxies is only 
about 3$\%$ to 10$\%$ of what is observed.  Further and perhaps even more significant, the 
energy spectrum of the Galactic diffuse $\gamma$-rays is significantly different 
from the extragalactic diffuse spectrum, being harder at low energies
($<$1000 MeV) and considerably steeper in the 1 to 50 GeV region
\cite{erlykin95}.  
Dar and Shaviv (1995)\cite{dar95} have suggested that emission arises from cosmic ray
interaction with intergalactic gas in groups and clusters of galaxies.  Although
the authors claim that this proposed explanation leads to a higher intensity level, 
it is still in marked disagreement with the measured energy spectrum \cite{stecker96a}.
Finally, cosmic ray electrons and protons that leak out into the
intergalactic space could upscatter the CMB to $\gamma$-ray energies.
Chi and Wolfendale (1989)\cite{chi89} and Wdowczyk and Wolfendale (1990)
\cite{ww90} have shown these contributions to be not significant.

\subsubsection*{Active Galaxies: Seyferts}

It has been postulated for over two decades by a large number of authors that active 
galactic nuclei (AGN) might be the source of the extragalactic $\gamma$-ray 
diffuse emission (e.g. \cite{bignami79}\cite{kazanas83}).  
However, prior to the launch of CGRO, only a handful of these objects were claimed to be detected at energies above 1 MeV.  With the detection of a large number of AGN with instruments on board CGRO, it was natural to  consider that the fainter unresolved sources could collectively make up the
observed extragalactic $\gamma$-ray emission. 
Using data from the OSSE experiment, it has been shown that the
average Seyfert galaxy spectrum, is
characterized by an exponentially falling continuum (e-folding energy
$\sim$100 keV) together with a Compton reflection component which
contributes mainly in the 10-50 keV band \cite{johnson94}\cite{zdziarski95}. 
However, none have been detected above 1 MeV \cite{maisack95}\cite{lin93}.
The generally accepted reason for this is that the photon fields within
AGN are so intense that pair production prevents the
higher energy $\gamma$-radiation from escaping \cite{svensson86}\
\cite{zdziarski86}\cite{done89}. Thus, although Seyfert galaxies may provide a
substantial contribution to the X-ray background, and may also provide the
dominant contribution to the high energy neutrino background \cite{stecker91}
\cite{stecker92}, these objects are not expected to be
important sources of high energy $\gamma$-radiation. Another sub-class of AGNs,
the `MeV-blazars', have been shown to exhibit
peak power at MeV energies \cite{bloemen95}\cite{blom95}.
Comastri, Girolamo and Setti\cite{cgs96} have discussed the possible contribution 
to the diffuse emission from these objects. 
Since at present, only $\sim$1$\%$ of the
$\gamma$-ray blazars show an MeV excess, observationally, it suggests that
MeV-blazars may not be a significant contributor to the extragalactic diffuse
emission.

\subsubsection*{Active Galaxies: Blazars}
While radio-quiet AGN do not qualify as significant
medium or high energy $\gamma$-ray
sources, the EGRET observations showed the presence of a class of
AGN, characterized by strong time variability at many wavelengths
including $\gamma$-rays, flat radio spectrum and often exhibit
strong polarization and (or) superluminal motion.
These are the BL Lac objects and the flat spectrum radio
quasars (FSRQ) which together have been classified as `blazars'. The 
EGRET team has now reported the detection of over 50 blazars 
\cite{cvm95}\cite{muk97}.
Of these six have been detected at COMPTEL energies as well. 
It is believed that most of these objects generally have
jets beaming their emission toward us. It is natural to
assume that, since the jets are optically thin to high energy
$\gamma$-rays and since beaming in the jets can produce a large 
enhancement in the apparent $\gamma$-ray luminosity relative to unbeamed 
components, therefore, the $\gamma$-ray emission from these objects is 
probably beamed and originates in the jets. This hypothesis is supported 
by the rapid $\gamma$-ray time variability observed in many blazars 
(for eg. 3C279 \cite{kniffen93}).

McNaron-Brown \etal (1995)\cite{mcnaron95} examined the multiwavelength spectra of the six blazars detected by
OSSE, COMPTEL and EGRET. Using this data set, one can derive an `average' 
blazar spectrum characterized by a broken power-law with a break at $\sim$4 MeV
(\cite{sreekumar97}).
The derived average differential photon spectral index below 4 MeV is
determined to be about --1.7 and above 4 MeV to be equal to the average
spectral index of EGRET detected blazars ($\sim$--2.1). 
However below 10 MeV, the blazar contribution alone is not
sufficient to explain the observed extragalactic emission. Thus, the
true origin of the emission in this energy range is not yet fully understood.
We note that while the EGRET spectrum represents the
extragalactic diffuse radiation, the COMPTEL spectrum refers to the
total $\gamma$-ray flux from high-latitude observations (including
contribution from the Galactic diffuse and $\gamma$-ray point sources
in the FOV of $\sim$1.5 sr).

One of the more important pieces of evidence in favor of the blazar origin 
of the high-energy portion of the diffuse spectrum is the spectrum itself.  
Both the spectrum reported here and the average spectrum of blazars may 
be well represented by a power law in photon energy.  The spectral index 
determined here for the diffuse radiation is --(2.10$\pm$0.03), and the 
average spectral index of the observed blazars is --(2.15$\pm$0.04) 
\cite{muk97} These two numbers are clearly in good agreement. A standard 
cosmological integration of a power law in energy yields the same 
functional form and slope. Considering the new, well determined $\gamma$-ray 
spectrum, this argues strongly that the bulk of the observed extragalactic 
$\gamma$-ray emission can be explained as ordinating from unresolved blazars.

In order to estimate the intensity of the diffuse radiation from blazars, 
knowledge of the evolution function is needed, as well as the intensity 
distribution of the blazars.  Two approaches have been utilized to determine 
the $\gamma$-ray evolution.  One way is to assume that the evolution is 
similar to that at other wavelengths.  The other alternative is to deduce 
the evolution from the $\gamma$-ray data itself and hence have a solution 
that depends only on the $\gamma$-ray results.  The advantage of the former 
is that, if the assumption of a common evolution is correct, an estimate 
with less uncertainty is obtained.  The positive aspect of the latter is 
that there is no assumption of this kind, but the uncertainty in the 
calculated results is relatively large because of the small $\gamma$-ray  
blazar sample.

Several authors have estimated the contribution from blazars using the 
first approach described above \cite{padovani93}\cite{stecker93}
\cite{setti94}\cite{erlykin95}\cite{stecker96b}\cite{kazanas97}
where one accepts the proposition that the evolution
function determined from the radio data may be applied to
the $\gamma$-ray-emitting blazars.
Furthermore, the recent work of Mukherjee \etal (1997)\cite{muk97} shows 
that there is general agreement within uncertainties between the radio 
and high-energy $\gamma$-ray redshift distributions of both types of 
blazars, i.e. flat-spectrum radio quasars and BL Lacs.
However, a word of caution seems appropriate, since it should be
pointed out that the degree and nature of the correlation between the radio and
$\gamma$-ray emission is still being debated \cite{padovani93}\cite{mucke97}
\cite{mattox97}.  Most of these calculations show that all or most of 
the observed mission can be explained as originating from unresolved blazars.
Stecker and Salamon (1996)\cite{stecker96b} instead of assuming a mean 
$\gamma$-ray spectral index, used the distribution of the observed spectral 
indices in their calculation. This introduces a curvature in the spectrum; 
the steep spectrum sources contributing more at lower energies ($<$500 MeV)
and the flat-spectrum sources dominating the emission at higher
energies.  This is consistent with the curvature in the spectrum reported by
Osborne, Wolfendale and Zhang (1994)\cite{osborne94}. However recent 
instrumental corrections to the 2-4 GeV energy range \cite{esposito97} and 
the extension of the spectrum to 100 GeV, have weakened any evidence for 
such a curvature in the spectrum.

Chiang \etal (1995)\cite{chiang95} used the second approach by  using the 
$\gamma$-ray blazar data to deduce the evolution function.  
They  used the V/Vmax approach in the context of pure luminosity evolution 
to show that there was indeed evolution of the high-energy $\gamma$-ray 
emitting blazars and found that the implied evolution is similar to that 
seen at other wavelengths.  
However, recent work of Chiang and Mukherjee (1997)\cite{chiang97} argue 
that an improved calculation of the lower end of the de-evolved
luminosity function indicates that only $\sim$25$\%$ of the observed
emission is made up of unresolved blazars. As stated before, the limited 
sample of detected $\gamma$-ray blazars results in larger 
uncertainty in the above calculation. Furthermore, Stecker and Salamon
(1996)\cite{stecker96b} and Kazanas and Perlman (1997)\cite{kazanas97} 
concluded that EGRET has preferentially detected those blazars that 
were in `flaring' states.  Thus there is a clear need
for a much improved evaluation of the true $\gamma$-ray luminosity
function. The expected detection of a large number of sources using a future 
more sensitive $\gamma$-ray instrument such as GLAST, could make this possible.

If the hypothesis that the general diffuse radiation is the sum of the emission of blazars is 
accepted, there is an interesting corollary.  The spectrum of the measured
extragalactic emission implies the average energy spectra of blazars extend 
to at least 50 GeV and maybe up to 100 GeV without a significant change in slope.  
Most of the measured spectra of 
individual blazars only extend to several GeV and none extend above 10 GeV, simply 
because the intensity is too weak to have a significant number of photons to
measure. Intergalactic absorption does not have much effect at this energy 
except for blazars at relatively large redshift, and, in any case would steepen
the spectrum at high energies.
Hence, the continuation of the single power law diffuse spectrum 
up to 100 GeV strongly suggests that the source spectrum also 
continues without a major change in spectral slope to at least 100 GeV.  
This conclusion, in turn, implies that the spectrum of the parent 
relativistic particles in blazars that produce the 
$\gamma$-rays remains hard to even higher energies.

\section*{Summary}

CGRO observations have lead to a significant advancement in our understanding 
of the extragalactic $\gamma$-ray emission. 
The recent COMPTEL measurements and SMM results have shown that a
significant part of the MeV-excess previously reported around 1--10 MeV, 
is due to instrumental background events. 
The new measurements in the 1--30 MeV range are compatible with power-law 
extrapolations from lower and higher energies.
The COMPTEL results on the 9--30 MeV flux represents
the first significant detection in this energy range. 
Above 30 MeV, the EGRET observations have extended the high-energy 
measurement to an unprecedented $\sim$100 GeV. 
The 30 MeV to 100 GeV spectrum is well described by a single
power-law with spectral index of -2.1. No large scale spatial
anisotropy or changes in the energy spectrum is observed in the
deduced extragalactic spectrum above 30 MeV.
The bulk of the extragalactic emission above 10 MeV appears to arise from
unresolved blazars and is supported by the
consistency in shape between the two spectra.
However, below 10 MeV, the exact nature of the emission
is not well understood, partly due to the large
uncertainties in the measured diffuse spectrum in this
energy range. The average blazar spectrum suggest that only  
about 50$\%$ of the measured emission in the 1--10 MeV range, could arise from 
blazars. Current observational limits do not provide tight constraints on 
contributions from additional source classes, or from other truly diffuse 
processes, making this an important area of investigation for 
the next generation $\gamma$-ray experiments. 

\acknowledgments
The authors wish to thank James Ryan (UNH) for helpful discussions.

\end{document}